\documentclass[conference]{IEEEtran}
\IEEEoverridecommandlockouts
\usepackage{cite}
\usepackage{amsmath,amssymb,amsfonts}
\usepackage{algorithm}
\usepackage{algpseudocode}
\algrenewcommand\algorithmicrequire{\textbf{Initialize:}}
\usepackage{graphicx}
\usepackage{textcomp}
\usepackage{xcolor}
\def\BibTeX{{\rm B\kern-.05em{\sc i\kern-.025em b}\kern-.08em
    T\kern-.1667em\lower.7ex\hbox{E}\kern-.125emX}}
\usepackage[
    top=0.71in,
    bottom=1.02in,
    left=0.7in,
    right=0.7in
]{geometry}

\begin{document}

\title{GPU-Accelerated Quantum Annealing-Inspired UAV Path Planning for Smart Agriculture
}

\author{Maho Hirahara\IEEEauthorrefmark{1},
Aohan Li\IEEEauthorrefmark{1}\IEEEauthorrefmark{2}
\\
\IEEEauthorrefmark{1}Department of Computer and Network Engineering, The University of Electro-Communications, Tokyo, Japan\\
\IEEEauthorrefmark{2}Meta-Networking Research Center, The University of Electro-Communications, Tokyo, Japan\\
}

\maketitle
\begin{abstract}
With the rapid advancement of smart agriculture, Unmanned Aerial Vehicles (UAVs) have become indispensable for efficient coverage and path planning in grid-mapped agricultural fields. However, optimizing UAV flight paths in large-scale environments remains a complex Nondeterministic Polynomial-time (NP)-hard problem. Traditional path planning methods are often constrained by local optima, limited scalability, and slow convergence, which significantly restrict their effectiveness in solving large-scale problems. To address these limitations, this paper proposes a novel optimization framework utilizing a Graphics Processing Unit (GPU)-parallelized Ising solver. Our method mimics the operational principles of quantum annealing on GPU hardware, enabling rapid search for the minimum-energy state of Ising models. Unlike physical quantum devices, which are often constrained by the number of qubits, our approach leverages the Fixstars Amplify (FA) platform to perform parallel annealing on highly parallelized GPUs. This enables the simultaneous evaluation of thousands of potential path candidates and vast state transitions. By leveraging large-scale parallel processing, this framework's core strength lies in minimizing computation time even as the problem scale increases. 
Simulation results demonstrate that our proposed method consistently identifies superior flight paths while maintaining stable computational performance compared with the genetic algorithm and simulated annealing method. 
\end{abstract}


\maketitle

\section{Introduction}
\label{Introduction}

Unmanned aerial vehicles (UAVs) have become indispensable in various fields due to their exceptional maneuverability, cost-effective deployment, and ability to be easily reprogrammed for diverse missions. For example, they are utilized for data collection and energy transfer support in rechargeable wireless sensor networks~\cite{b1}, providing efficient coverage of complex 3D structures such as wind turbines and bridges~\cite{b2}, and maximizing total transmission rates in multi-UAV-assisted wireless communication networks~\cite{b3}. In particular, for pre-disaster assessments, UAV-based path planning is essential for generating optimal, collision-free trajectories in dense or maze-like obstacle environments to identify the causes of disasters such as wildfires~\cite{b4}.

In light of these advancements, strategic UAV integration is also progressing in smart agriculture for tasks such as pesticide spraying and real-time crop monitoring \cite{b5}. Recent research has developed various meta-heuristic algorithms to navigate the search space in the agricultural sector, such as cooperative path planning for heterogeneous UAV swarms \cite{b6} and hybrid particle swarm optimization (PSO) techniques \cite{b7}. These methods aim to improve operational efficiency and reduce labor intensity compared to conventional manual approaches. However, existing UAV-based approaches in agriculture still face significant practical constraints, such as limited battery life, low payload capacity, and difficulties in efficiently covering large areas \cite{b5}.

Furthermore, the use of grid-based environmental modeling has been widely adopted to manage mission planning and task allocation in UAV networks~\cite{b8, b9}. For instance, hierarchical discrete grid structures have been employed for conflict detection and path planning~\cite{b8}, while distributed online cooperation methods utilizing environmental information maps have been developed for coverage path planning to minimize task completion time~\cite{b9}. However, in these grid-based cooperative planning scenarios that involve multiple conflicting objectives, the optimization models are inherently Non-deterministic Polynomial-time (NP)-hard. As the scale of the farmland expands, the exponential increase in the number of grid cells leads to a combinatorial explosion of the search space. Consequently, conventional approaches suffer from prohibitive computational overhead, making it difficult to achieve real-time path generation or to maintain operational efficiency in large-scale agricultural environments where battery life and mission time are strictly limited~\cite{b10}.

To address these issues, quantum annealing (QA), which rapidly solves optimization problems based on quantum-inspired physics, has gathered attention~\cite{b3, b11, b12, b13}. QA transforms combinatorial optimization problems into quadratic unconstrained binary optimization (QUBO) models. By searching for the physical minimum energy state, it efficiently identifies solutions in vast search spaces that are challenging for conventional methods. However, applying current physical quantum computing technology directly to large-scale agricultural settings remains premature due to challenges regarding hardware scale and stability \cite{b14}. To overcome this challenge and realize the superior search capabilities of QA using realistic computational resources, this study focuses on the parallelism of computational architectures based on the Ising model. The Fixstars Amplify platform is a cloud environment dedicated to Ising computations and is used to map this QUBO model onto highly parallelized Graphics Processing Units (GPUs) \cite{b15}. This process finds the minimum energy state of the Ising model, effectively mimicking the operating principles of quantum annealing. Unlike physical quantum devices, this approach achieves this through parallel annealing on GPUs. By simulating multiple candidate paths simultaneously, it enables the parallel evaluation of vast numbers of state transitions, allowing for rapid exploration of complex solution spaces.

Motivated by the above discussion, we propose a GPU-accelerated quantum annealing-inspired UAV path planning method for smart agriculture, enabling real-time and scalable agricultural spraying in this paper. In the proposed method, the path planning problem is first formulated as a QUBO model. In the QUBO model, the concept of energy states from physics is utilized as an optimization metric. Specifically, evaluation metrics such as total flight time are quantified and expressed as a single energy value integrated with the constraints. A lower value indicates a superior solution that satisfies the constraints while achieving the objective. Through this mathematical representation, various constraints and objective functions are expressed as energy states, enabling the solver to identify a global minimum solution within the vast search space.

The main contributions of this paper are summarized as follows:
\begin{itemize}
\item We propose a novel GPU-accelerated QA-inspired path optimization framework for real-time and scalable agricultural spraying. The main advantage of the proposed method lies in its ability to explore the solution space through large-scale parallel processing rather than iterative sequential updates. Physical quantum processing units (QPUs) offer the theoretical benefits of QA, but current limitations in the number of qubits and connectivity make it difficult to achieve practical applications. In contrast, by adopting a GPU-parallelized Ising solver, we can leverage the high scalability of digital circuits to process tens of thousands of variables simultaneously.
\item We formulated the UAV path planning problem in precision agriculture as a QUBO model for solving it using a GPU-parallelized Ising solver, incorporating agricultural-specific constraints. In the formulated optimization problem, we consider the minimization of the time for spraying under the constraints that each grid is visited exactly once, and the UAV is located at exactly one grid at each time.
\item We implemented and evaluated the optimization using the GPU-parallelized Ising solver (Fixstars Amplify), demonstrating its scalability for large-scale farm spraying. Simulation results show that the proposed method achieves superior performance compared to conventional metaheuristics---specifically Genetic Algorithm (GA) and Simulated Annealing (SA)---in balancing the mission completion time and strict adherence to coverage constraints.
\end{itemize}

This paper is structured as follows. Section \ref{System Model} describes the system model and problem formulation. Section \ref{Proposed Method} presents the proposed method and Ising mapping. Section \ref{Performance Evaluation} provides the implementation and performance evaluation. Finally, Section \ref{Conclusion} concludes the paper.

\section{System Model}
\label{System Model}

\begin{table}[t]
\caption{Summary of Notations}
\label{tab:Summary of Notations}
\begin{center}
\resizebox{\linewidth}{!}{
\begin{tabular}{|c|c|}
\hline
\textbf{Symbol} & \textbf{Description} \\
\hline
$M, N$ & Number of grid cells in row and column\\
\hline
$(i,j)$ & Grid coordinates within agricultural land \\
\hline
$w_{grid}$ & Length of a single square grid cell \\
\hline
$\tau$ & The flight time per unit distance \\
\hline
$n$ & Total number of time steps ($n = M \times N$) \\
\hline
$x_{t,i}$ & Binary variable; 1 if UAV is at cell $i$ at step $t$, 0 otherwise \\
\hline
$d_{i,j}$ & Euclidean distance between cell $i$ and cell $j$ \\
\hline
$t_{i,j}$ & Travel time between cell $i$ and cell $j$ \\
\hline
$\mathbf{T}$ & $n \times n$ travel time matrix consisting of $t_{i,j}$ \\
\hline
$H_{dist}$ & Objective function term for minimizing total travel distance \\
\hline
$H_{visit}$ & Penalty term ensuring each grid cell is visited exactly once \\
\hline
$H_{time}$ & Penalty term ensuring the UAV is at only one cell per time step \\
\hline
$A,B$ & Penalty weights for constraint satisfaction \\
\hline
$L$ & Computation time limit for the Ising solver (ms) \\
\hline
$P$ & UAV flight path (sequence of locations)\\
\hline
\end{tabular}
}
\label{tab:param_settings}
\end{center}
\end{table}

\begin{figure}[t]
\centering
\includegraphics[width=0.80\linewidth]{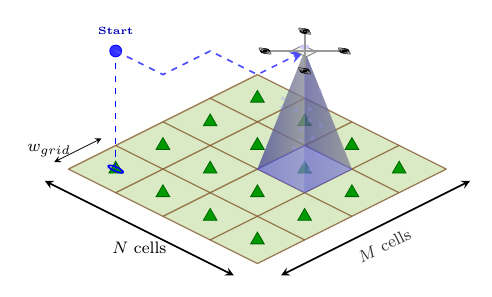}
\caption{System Model}
\label{fig:s1}
\end{figure}

In this study, the target agricultural field is discretized into a grid-based environment to facilitate path planning. The field is divided into $M \times N$ rectangular cells, where each cell represents a basic unit of the task area. Each grid cell is a square with a side length of $w_{grid}$, resulting in a total area of $M \times N \times w_{grid}^2$. Let $c_{i,j}$ denote the cell located at the $i$-th row and $j$-th column, where $0 \le i \le M-1$ and $0 \le j \le N-1$. The center of each cell is considered a point that the UAV must navigate through. The UAV system examined in this study is shown in Fig.~\ref{fig:s1}. Table \ref{tab:Summary of Notations} summarizes the notations used throughout the paper.

Based on these coordinates, an $n \times n$ travel time matrix $\mathbf{T}$ is initialized to define the travel costs between all possible pairs of locations. Since the UAV must visit every cell in the field, both the total number of locations and the required number of time steps are set to $n = M \times N$. This matrix $\mathbf{T}$ serves as the matrix of quadratic coefficients for the objective function used to determine the optimal flight path $P$. Each element $t_{i,j}$ represents the travel time from a departure cell $i$ to an arrival cell $j$, calculated as $t_{i,j} = d_{i,j} \times \tau$. Here, $d_{i,j}$ is the Euclidean distance between these two cells, and $\tau$ is the flight time per unit distance. In the optimization process, we define a binary decision variable $x_{t,i}$, which takes the value $1$ if the UAV is located in cell $i$ at time step $t$, and $0$ otherwise. To ensure a valid tour, the problem is formulated into a QUBO structure by defining constraint functions, which apply one-hot constraints across both the spatial and temporal axes. These constraints guarantee that each grid cell is visited exactly once and that the UAV occupies only one cell at each time step. Furthermore, a fixed starting point constraint is set as $x_{0,0} = 1$.

The primary objective is to find the optimal sequence of grid visits that minimizes the total flight time of the UAV while ensuring that all target cells are covered. To formally define the optimization problem, let ${K} = \{0, \dots, n-1\}$ be the set of discrete time steps and $G = \{0, \dots, n-1\}$ be the set of grid cell indices. The optimization problem is formulated as follows:

\begin{subequations}
\begin{align}
(\textbf{P1}) &\min_{x_{t,i}} \sum_{t=0}^{n-1} \sum_{i=0}^{n-1} \sum_{j=0}^{n-1} t_{i,j} \cdot x_{t,i} \cdot x_{t+1,j}\\
{\rm s.t.} \quad 
&C1: \sum_{i=0}^{n-1} x_{t,i} = 1, \quad \forall t \in K   \\
&C2: \sum_{t=0}^{n-1} x_{t,i} = 1, \quad \forall i \in G \\
&C3: x_{t,i} \in \{0, 1\}. 
\end{align}
\end{subequations}

Constraint C1 ensures that the UAV is located at exactly one cell at each time step $t$. Constraint C2 guarantees that each cell $i$ is visited exactly once throughout the entire mission, satisfying the coverage requirement. Constraint C3 enforces the binary constraint on $x_{t,i}$.

\section{Proposed Method}
\label{Proposed Method}
In this section, we describe the proposed GPU-accelerated quantum annealing-inspired UAV path planning method. The core idea is to transform the UAV path planning problem into a QUBO form and solve it using high-speed parallel GPU processing. The remainder of this section is organized as follows. First, QA is introduced as the theoretical foundation using the Ising model. Next, we present the QUBO formulation for UAV path planning, detailing the mathematical mapping of flight objectives and physical constraints. Finally, the proposed algorithm is described, covering the procedural flow from initialization to GPU-based execution and path reconstruction.

\subsection{Quantum Annealing}
QA is a computational framework that utilizes the principles of quantum mechanics to solve combinatorial optimization problems. Its basic mechanism involves controlling a spin system to search for the minimum-energy state of the Ising model. The spin system is defined by the Ising variables $\sigma_k \in \{+1, -1\}$. Here, $k$ represents the index (subscript) when the decision variables defined in multiple dimensions are rearranged into a one-dimensional array. The Hamiltonian, which is the energy functional, is expressed as follows~\cite{b11}:
\begin{equation}
H_{Ising}(\sigma) = \sum_{k=1}^{S} h_k \sigma_k + \sum_{k<l} J_{kl} \sigma_k \sigma_l
\end{equation}

In this equation, $J_{kl}$ represents the interaction coefficient between the $k$-th and $l$-th spins, and $J_{kl} \sigma_k \sigma_l$ represents the interaction energy between the two Ising variables $\sigma_k$ and $\sigma_l$. Furthermore, $S$ denotes the number of Ising spins, $h_k$ is the external magnetic field of the $k$-th spin, and $h_k \sigma_k$ represents the energy contribution of the $k$-th spin. In this study, we map the constraints and objective function of UAV path planning onto this Ising structure and derive the optimal solution to the combinatorial optimization problem by minimizing the resulting Hamiltonian.

\subsection{QUBO Formulation for UAV Path Planning}
The formulated problem is solved using a GPU-accelerated quantum annealing-inspired solver. Mission parameters, including the grid dimensions ($M, N, w_{grid}$) and constraints, are converted into QUBO format and fed into the GPU-based solver. Here, we detail how to formulate the UAV path planning problem as a QUBO model.

The objective is to find the optimal sequence of grid visits that minimizes the total flight time of the UAV while ensuring that all target cells are covered. The objective function (1a) can be transformed into an energy function, which is expressed as follows:
\begin{equation}
H_{dist} = \sum_{t=0}^{n-1} \sum_{i=0}^{n-1} \sum_{j=0}^{n-1} t_{i,j} \cdot x_{t,i} \cdot x_{t+1,j}.
\end{equation}
In addition, the constraints (1b) and (1c) can be formulated as constraint terms as follows:
\begin{equation}
H_{visit} = \sum_{i=0}^{n-1} \left( 1 - \sum_{t=0}^{n-1} x_{t,i} \right)^2,
\end{equation}
\begin{equation}
H_{time} = \sum_{t=0}^{n-1} \left( 1 - \sum_{i=0}^{n-1} x_{t,i} \right)^2.
\end{equation}
Then, the final QUBO formulation (Hamiltonian) to be minimized by the Ising machine is expressed as follows:
\begin{equation}
H_{total} = AH_{dist} +\lambda_p (H_{visit} + H_{time}) 
\end{equation}
where $A$ and $\lambda_p$ are parameters that balance the importance of the objective and constraints to more likely reach the optimal solution. The ultimate goal is to minimize the total Hamiltonian $H_{total}$. The mission consists of a sequence of $n$ time steps, where $n = M \times N$ corresponds to the total number of grid cells to be covered.

\subsection{GPU-Accelerated QA Solver}
In optimization computing, Ising-based hardware solvers often demand rigid formatting and hardware-specific requirements to process mathematical equations. The primary role of the GPU-accelerated QA solver is to simplify these complex requirements, operating as an intermediate interface that automatically translates a user's abstract optimization problem into a machine-executable format.

This conversion workflow involves several automated pre-processing steps handled entirely by the Software Development Kit (SDK). When an optimization model consisting of objective functions and constraints is input, the SDK analyzes the equations to automatically construct penalty functions for constraints, reduce higher-order polynomial terms into quadratic forms, and map variables onto the actual calculation grid of the solver. This automation eliminates the need for manual algebraic adjustment, allowing direct deployment of the mathematical model to the backend engine.

The Amplify Annealing Engine (AE) is one of the high-performance QA-based optimization solvers developed by Fixstars for solving combinatorial optimization problems. It efficiently solves optimization problems formulated as QUBO and can be accessed through a unified interface provided by the Fixstars Amplify SDK. The actual optimization is performed by the Amplify AE, a cloud-based accelerator designed specifically for high-performance parallel computing. Unlike standard calculation systems, the Amplify AE utilizes massive GPU concurrency to run multiple annealing threads in parallel. This mechanism allows the engine to explore a vast space of path candidates simultaneously. By checking numerous states at the same time, the solver effectively avoids getting trapped in shallow, suboptimal solutions, maintaining a stable and efficient execution time even as the grid size of the agricultural field expands.

\subsection{UAV Path Planning Using GPU-Accelerated QA Solver}
The computation pipeline of the proposed method is managed by the Fixstars Amplify SDK and executed via the Amplify AE. Amplify AE solves the formulated optimization problem and determines the optimal flight path for the UAV. The detailed process is summarized in Algorithm 1.

The system first initializes the environment by computing the travel time matrix $T_{i,j}$ from grid coordinates defined by the $M \times N$ grid dimensions and the cell width $w_{grid}$ (line 1). To mathematically represent the trajectory, a binary decision variable $x_{t,i}$ is defined for each time step $t$ and cell $i$ (lines 2 $\sim$ 6), indicating whether the UAV occupies a specific position at a specific time. Next, the constraint function $H_{constraints}$ is formulated (line 7). This incorporates the visit requirement $H_{visit}$ for a specific cell and the temporal occupancy constraint $H_{time}$. This function $H_{constraints}$ is designed to reach a minimum state with zero energy only when all specified flight conditions are fully satisfied. Simultaneously, an objective function $H_{dist}$ is constructed to evaluate and minimize the cumulative travel time over successive steps (line 8). These components are integrated into a single unified formulation, resulting in a quadratic unconstrained binary optimization (QUBO) model denoted as $H_{QUBO}$ (line 9). In this formulation, parameters $A$ and $\lambda_p$ are introduced to control the balance between constraint satisfaction and cost minimization.

Next, during the optimization phase, the completed model is sent to the engine along with the time limit $L$ of the specified solver (line 10). The Amplify AE sampler is executed to explore the vast configuration space through large-scale parallel processing and identify the global minimum energy state of $H_{QUBO}$ (line 11). Upon completion, the solver extracts the optimal spin combination $x^*_{t,i}$. Finally, in the output reconstruction phase, the time-series flight trajectory $P$ and the minimum mission cost are reconstructed from the evaluated variables $x^*_{t,i}$, generating the final feasible flight path (line 12).

\begin{algorithm}[t]
\caption{UAV Path Planning via QA Solver}
\label{alg:amplify_detailed}
\begin{algorithmic}[1]
\Statex \textbf{Input:} Grid dimensions $M \times N$, cell width $w_{grid}$, unit flight time $\tau$, and solver time limit $L$.
\Statex \textbf{Output:} The optimal flight path $P$ and the total flight time.
\State Calculate travel time matrix $T_{i,j}$ based on grid coordinates
\For{$t = 0$ \textbf{to} $n-1$}
    \For{$i = 0$ \textbf{to} $n-1$}
        \State Define the binary variable $x_{t,i}$
    \EndFor
\EndFor
\State $H_{constraints} \leftarrow H_{visit} + H_{time}$
\State $H_{dist} \leftarrow \sum_{t=0}^{n-2} \sum_{i=0}^{n-1} \sum_{j=0}^{n-1} t_{i,j} \cdot x_{t,i} \cdot x_{t+1,j}$
\State $H_{QUBO} \leftarrow AH_{dist} + \lambda_p H_{constraints}$
\State Import $H_{QUBO}$ to Amplify AE sampler
\State Run the Amplify AE sampler with time limit $L$
\State Return the optimal spin combination $x^*_{t,i}$ and $P$
\end{algorithmic}
\end{algorithm}

\section{Performance Evaluation}
\label{Performance Evaluation}

In this section, we first introduce the GPU-accelerated annealing process and the baseline methods for comparison. Then, we present the simulation results for the proposed method in terms of total flight time and computation time compared with the baseline methods.

\subsection{GPU-Accelerated Annealing Process}
The proposed method is implemented using Fixstars Amplify, a platform that combines a unified SDK with a high-performance GPU-based Annealing Engine. While hardware-based quantum annealers often face scalability limits, this engine leverages massive GPU parallelism to solve complex combinatorial optimization problems. Unlike traditional sequential simulated annealing, the Amplify Engine executes parallel annealing, simulating thousands of independent state replicas simultaneously. This architecture enables high-density exploration of the solution space, significantly surpassing the capabilities of CPU-based processing. All simulations and baseline evaluations (GA and SA) were conducted on an Apple Mac mini equipped with an Apple M4 chip (10-core CPU, 10-core GPU) and 24 GB of unified memory.

\subsection{Baseline Methods for Comparison}
To benchmark the performance of the proposed GPU-accelerated QA-based method, we compare it against two widely used metaheuristic algorithms.
\begin{itemize}
    \item SA \cite{b16}: A probabilistic technique for approximating the global optimum of a given function. In this study, we use a standard CPU-based SA to solve the same QUBO formulation.
    \item GA \cite{b17}: A search heuristic inspired by the process of natural selection. GA is commonly used for TSP-like problems, employing crossover and mutation operators to evolve a population of potential paths.
    \item Exact method (brute force): An exhaustive search method that evaluates all permutations to find a globally optimal solution, and is used to determine the theoretical true value for small-scale problems.
\end{itemize}

\subsection{Simulation Parameters}
To evaluate the effectiveness of the proposed method, we conducted numerical simulations using the Fixstars Amplify platform. The agricultural field is modeled as an $M \times N$ grid with a unit width $w_{grid} = 5.0$ m, and the travel speed of the UAV is set to $5.0$ m/s. The weight coefficients for the objective function and constraints, i.e., $A$ and $\lambda_p$, were set to 1.
To ensure real-time feasibility in dynamic agricultural environments, the computation time limit $L$ for optimization tasks in the Amplify Annealing Engine (AE) was set to 2000 ms. Problem scale was varied across distinct grid dimensions ($3 \times 3$, $3 \times 4$, $4 \times 4$, $4 \times 5$, $5 \times 5$, $5 \times 6$, $6 \times 6$; 9--36 nodes) to analyze the scalability of each method.
For SA, the number of annealing samples was set to 1000, the random seed for reproducibility to 25, and the annealing schedule was controlled. For the GA, to maintain genetic diversity during the search process, the population size was set to 512, the number of parents for crossover to 10, the mutation probability to 0.2, the adjusted power coefficient to 1.0, and the number of generations to execute to 1000. The simulation environment and parameters are summarized in Table~\ref{tab:param_settings}.

\begin{table}[t]
\caption{Experimental Parameter Settings}
\begin{center}
\begin{tabular}{|c|c|}
\hline
\textbf{Parameter} & \textbf{Value} \\
\hline
Grid size & 3 × 3, 3 × 4, 4 × 4\\
& 4 × 5, 5 × 5, 5 × 6, 6 × 6 \\
\hline
Grid width & 5.0 m \\
\hline
UAV travel speed & 5.0 m/s \\
\hline
Weight coefficients $A$, $\lambda_p$ & 1.0\\
\hline
Computation time limit $L$ (Amplify AE) & 2000 ms \\
\hline
Penalty weight for constraints $\lambda_p$ (Amplify AE)& 1.0 \\
\hline
Number of annealing samples (SA) & 1000 \\
\hline
Random seed for reproducibility (SA) & 25 \\
\hline
Population size (GA) & 512 \\
\hline
Number of parents for crossover (GA) & 10 \\
\hline
Mutation probability (GA) & 0.2 \\
\hline
Adjusted power coefficient (GA) & 1.0 \\
\hline
Number of generations to execute (GA) & 1000 \\
\hline
\end{tabular}
\label{tab:param_settings}
\end{center}
\end{table}

\subsection{Results}

\begin{figure}[t]
\centering
\includegraphics[width=0.92\linewidth]{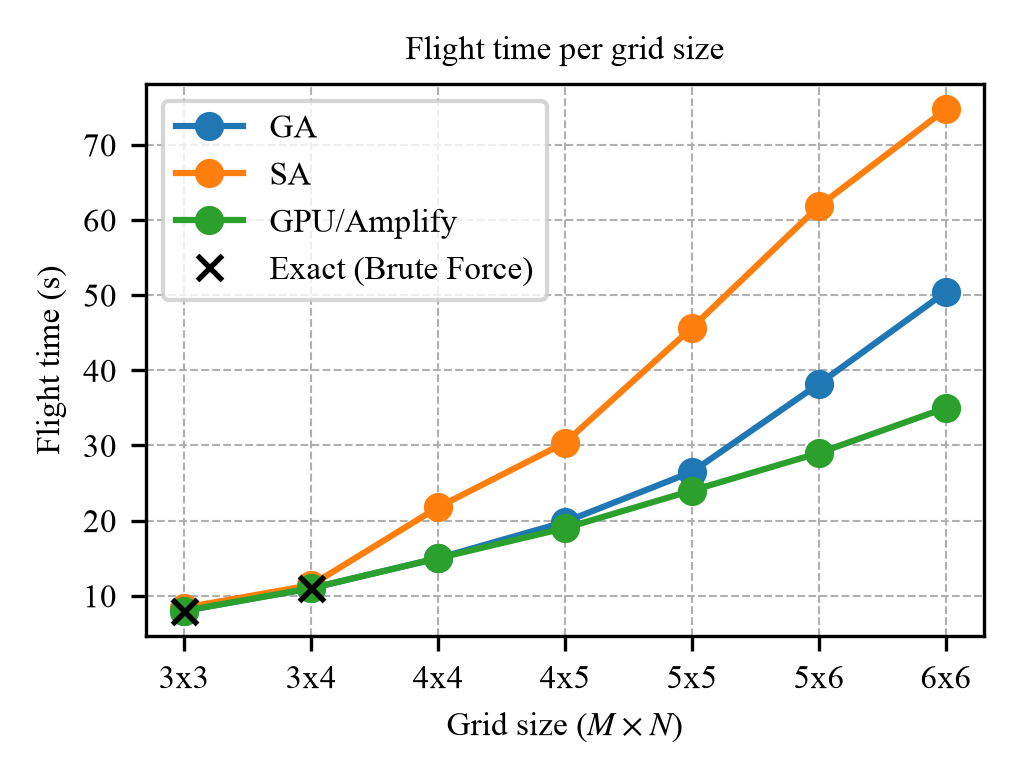}
\caption{Flight Time vs. Grid Size}
\label{Flight time}
\end{figure}

Simulations were conducted at various grid sizes ranging from $3 \times 3$ to $6 \times 6$ to evaluate the impact of increasing node density on computational efficiency.

\subsubsection{Flight time vs. grid size} 
Figure \ref{Flight time} shows the total flight time of the UAV for each grid size. When the grid size exceeded $3 \times 4$, it became impossible to perform a brute-force search due to memory constraints caused by the combinatorial explosion.
Figure \ref{Flight time} shows that the proposed method outperforms both GA and SA for all grid sizes. Furthermore, for grids up to $3 \times 4$, the proposed method produces the same solution as the optimal solution obtained through exhaustive search. Additionally, it is clearly demonstrated that as the problem size increases, the proposed method yields solutions of superior quality.
Specifically, on small grids like $3 \times 3$ or $3 \times 4$, the flight times of the three methods are nearly equivalent, suggesting that conventional heuristics can find near-optimal paths because the search space is sufficiently small. However, as the grid expands to 4×5, 5×6, and 6×6, a clear performance difference emerges. While SA and GA exhibit steeper cost increases and tend to settle for suboptimal routes in complex environments, the proposed method on the Fixstars Amplify platform consistently identifies paths with the minimum total distance. The difference in flight time becomes most pronounced on the $6 \times 6$ grid, demonstrating the robustness of the proposed method for large-scale path planning.

Compared to SA and GA, the proposed method can output an optimized solution unaffected by increases in the number of nodes. The superior solution quality of the proposed method stems from its ability to conduct a broader exploration of the energy space within the same time frame. The parallel annealing mechanism allows the solver to escape local optima more effectively than traditional SA, which often gets trapped in sub-optimal paths as the problem's constraints become more intricate. By maintaining high exploration density across a large-scale grid, the proposed method consistently identifies more efficient traversal routes, thereby minimizing the total flight time for pesticide spraying.

\begin{figure}[t]
\centering
\includegraphics[width=0.92\linewidth]{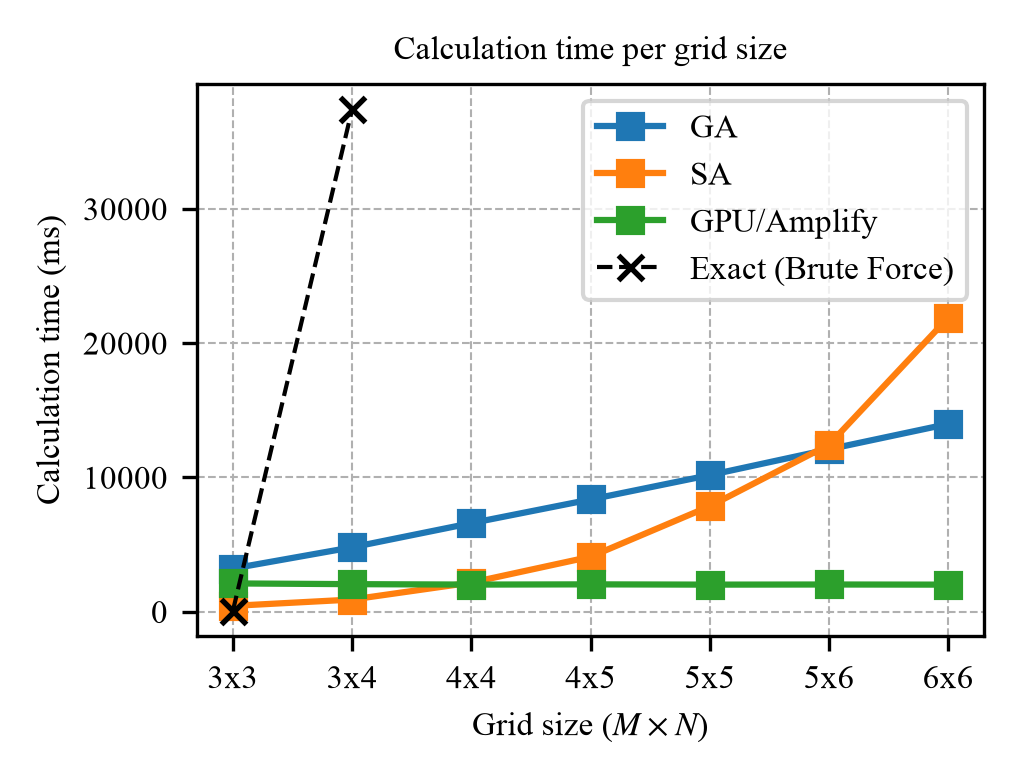}
\caption{Calculation Time vs. Grid Size}
\label{Calculation time}
\end{figure}

\subsubsection{Calculation Time Stability}
Fig. \ref{Calculation time} shows the computation time for each grid size. As shown in Figure \ref{Calculation time}, while SA and GA typically exhibit a significant increase in execution time as the number of nodes increases, the proposed method demonstrates computational stability. Additionally, the brute-force search ran out of memory starting at $4 \times 4$.
Specifically, GA required the most computational time for all grid sizes. For small configurations like $3 \times 3$ and $3 \times 4$, SA outperformed Fixstars Amplify, achieving the lowest computational time. However, when the grid exceeded $4 \times 4$, SA's execution time rose sharply, reflecting the exponential growth of the problem's search space. In contrast, the proposed method maintained a nearly flat line on the graph, indicating that processing time was largely independent of grid size within the tested range. Even on the largest $6 \times 6$ grid, where complexity peaks, Fixstars Amplify consistently returns solutions in a fraction of the time required by conventional metaheuristic methods.
This stability is attributed to the massive parallel processing architecture of the GPU-based Ising solver. While conventional metaheuristics like SA and GA process potential solutions or generations sequentially, the proposed method evaluates numerous state transitions simultaneously. Consequently, even as the search space expands with the number of nodes, the parallelized annealing process can effectively manage the increased complexity without a proportional increase in physical computation time.

\section{Conclusion}
\label{Conclusion}
In this paper, we proposed a GPU-accelerated quantum annealing-inspired path planning method for smart agriculture with UAVs, leveraging the parallel processing capabilities of the Fixstars Amplify GPU Ising Machine. By formulating the grid-based motion problem within a QUBO framework, we demonstrated the ability to solve NP-hard combinatorial optimization problems associated with expanding farmland with high stability. Performance evaluation results for grid dimensions ranging from $3\times3$ to $6\times6$ revealed that the proposed method can consistently identify optimal paths in less than 320 milliseconds. This efficiency is critical for real-time UAV operations that require immediate path updates due to battery constraints and changing field conditions. This initial research focused on a simplified flight time-based Hamiltonian to verify the performance of the basic solver. The next step is to integrate physical factors such as battery capacity and pesticide application amount into the optimization objectives and aim for computation and evaluation on larger $M \times N$ grids.

\end{document}